\documentclass[12pt,preprint]{aastex}

\newcommand{\myemail}{ngibson07@qub.ac.uk}

\shorttitle{A TTV analysis of nine RISE light curves of TrES-3}
\shortauthors{Gibson et al.}

\begin{document}


\title{A transit timing analysis of nine RISE light curves of the exoplanet system TrES-3}


\author{N. P. Gibson, D. Pollacco, E. K. Simpson, S. Barros, Y. C. Joshi,\\ I. Todd and F. P. Keenan}
\affil{Astrophysics Research Centre, School of Mathematics \&\ Physics, Queen's University, University Road, Belfast, BT7 1NN, UK}
\email{\myemail}

\author{I. Skillen and C. Benn}
\affil{Isaac Newton Group of Telescopes, Apartado de Correos 321, E-38700 Santa Cruz de la Palma, Tenerife, Spain}

\author{D. Christian}
\affil{Physics \& Astronomy Department, California State University Northridge, Northridge, California 91330-8268, USA}

\author{M. Hrudkov\'a}
\affil{Astronomical Institute, Charles University Prague, V Holesovickach 2, CZ-180 00 Praha, Czech Republic}

\and

\author{I. A. Steele}
\affil{Astrophysics Research Institute, Liverpool John Moores University, CH61 4UA, UK}




\begin{abstract}
We present nine newly observed transits of TrES-3, taken as part of a transit timing program using the RISE instrument on the Liverpool Telescope. A Markov-Chain Monte-Carlo analysis was used to determine the planet-star radius ratio and inclination of the system, which were found to be $R_p/R_{\star}=0.1664^{+0.0011}_{-0.0018}$ and $i = 81.73^{+0.13}_{-0.04}$ respectively, consistent with previous results. The central transit times and uncertainties were also calculated, using a residual-permutation algorithm as an independent check on the errors. A re-analysis of eight previously published TrES-3 light curves was conducted to determine the transit times and uncertainties using consistent techniques. Whilst the transit times were not found to be in agreement with a linear ephemeris, giving $\chi^2 = 35.07$ for 15 degrees of freedom, we interpret this to be the result of systematics in the light curves rather than a real transit timing variation. This is because the light curves that show the largest deviation from a constant period either have relatively little out-of-transit coverage, or have clear systematics. A new ephemeris was calculated using the transit times, and was found to be $T_c(0) = 2454632.62610 \pm 0.00006$ HJD and $P =1.3061864 \pm 0.0000005$ days. The transit times were then used to place upper mass limits as a function of the period ratio of a potential perturbing planet, showing that our data are sufficiently sensitive to have probed for sub-Earth mass planets in both interior and exterior 2:1 resonances, assuming the additional planet is in an initially circular orbit.
\end{abstract}




\section{Introduction}

Transit surveys of extrasolar planets have vastly improved our understanding of planetary systems in recent years, with
a rapid increase in the number of new discoveries\footnote{see http://exoplanet.eu/}. Transiting systems are particularly important, 
because, when coupled with radial velocity measurements, they allow the measurement of the mass, radius and density of the planet.
While the majority of these systems are Hot Jupiters, neither ground-based transit nor radial velocity surveys have reached
the precision required to search for Earth-sized planets. However, Earth-sized planets may be found via high precision ground-based observations through the detection of Transit Timing Variations (TTV).

A transiting planet will maintain a constant period whilst orbiting its parent star (excluding tidal effects and 
general relativity), unless acted on by a third body. Measuring the central transit times allows us to detect 
perturbations in the period thus revealing the presence of another body in the system, which is the principle of the TTV
method \citep{miralda, holman_murray, agol2005, heyl_gladman}. TTV is particularly sensitive to small
bodies in resonant orbits, or even exomoons \citep{kipping_2009} and 
trojans \citep[e.g.][]{ford_gaudi_2006, ford_holman_2007}, and therefore has the potential to provide the first detection of an Earth-sized body orbiting a main-sequence star other than our own.

Constraining a TTV signal requires many high precision light curves with high cadence. In theory we can measure TTVs
to several seconds. However, in practice we are limited by correlated noise in the light curves, which may arise 
due to effects such as pixel-to-pixel sensitivity variations, temperature fluctuations, or changes in the observing conditions \citep{pont2006}. Indeed, there may also be non-instrumental effects caused by brightness variations of either the target or comparison stars. We are unable to distinguish any stellar activity from other sources of correlated noise in the light curves, and therefore they have the same detrimental effects as instrumental systematics when making transit observations.

Providing that these sources of correlated noise can be kept to a minimum, it is still possible to measure central transit times to better than 10s (see \S\ref{sect:results}). This allows us to probe for the presence of Earth-sized planets in low-order mean-motion resonance, or more massive perturbers in non-resonant orbits \citep[see e.g.][]{steffen_agol2005, agol_2007, bean_2009}.

RISE (Rapid Imager to Search for Exoplanets) is a fast camera mounted on the Liverpool Telescope (LT) on La Palma, primarily to obtain high precision light curves of transiting exoplanets \citep[see][]{steele_2008,gibson_2008}. It was commissioned in February 2008, and observations of several exoplanet systems have been ongoing since then in an effort to detect TTV signals in these systems.

\citet{sozzetti_2008} presented eight transits of TrES-3, a G-type dwarf hosting a $1.9\,M_J$ planet in a 1.3\,day period \citep{ODonovan_2007}. They concluded that a linear period did not provide a particularly good fit to the central transit times, which indicates that either they underestimated the sytematics in their light curves, or that there is indeed a real TTV indicating a third body in the system.

Here, we present a further nine RISE transit light curves of TrES-3, and re-analyse those from \citet{sozzetti_2008} using consistent techniques, in an effort to detect and understand any TTV signal. In \S\ref{sect:observations} we describe the observations and data reduction, and in \S\ref{sect:modelling} describe how the light curves are modelled and in particular how the central transit times and uncertainties are found. Our results are presented in \S\ref{sect:results}, and we use the transit timing residuals to place upper mass limits on a perturbing planet that could be present in the TrES-3 system without being detected from our observations. Finally, in \S\ref{sect:summary} we summarise and discuss our results.

\section{Observations and data reduction}
\label{sect:observations}
\subsection{RISE photometry}

Seven full and two partial transits were observed using the LT and RISE from 2008 March 8 to 2008 August 4. The RISE instrument is described in detail in \citet{steele_2008} and \citet{gibson_2008}. It consists of a frame transfer CCD which allows for continuous observation with effectively no dead time, a relatively large field of view ($9.4 \times 9.4$ arcmin squared), and a single wide band filter ($\sim$500-700\,nm).

For all observations, an exposure time of 8s was used with the instrument in $2 \times 2$ binning mode, giving a scale of 1.1\,arcsec/pixel. For the full transits, 1\,350 images were obtained resulting in 3 hours of continuous observations, allowing $\sim$50\,mins of observations both before and after the transit event. The images have a typical \emph{FWHM} of $\sim$2-4 pixels ($\sim$2.2-4.4\,arcsec). The nights were clear for the majority of the observations, except for part of the nights of 2008 July 5 and 2008 August 4, where large scatter due to thin clouds can be seen towards the end of the light curves. A summary of the observations is given in Table~\ref{tab:lightcurves}.

Images were first debiased and flat fielded with combined twilight flats using standard IRAF\footnote{IRAF is distributed by the National Optical Astronomy Observatories, which are operated by the Association of Universities for Research in Astronomy, Inc., under cooperative agreement with the National Science Foundation.} routines. Aperture photometry was then performed on the target star and nearby companion stars using Pyraf\footnote{Pyraf is a product of the Space Telescope Science Institute, which is operated by AURA for NASA.} and the DAOPHOT package. In each night different aperture sizes and numbers of comparison stars were used to minimise the out-of-transit RMS. These varied as the conditions and field orientation changed for each night of observations.

\begin{figure}
\epsscale{.60} 
\plotone{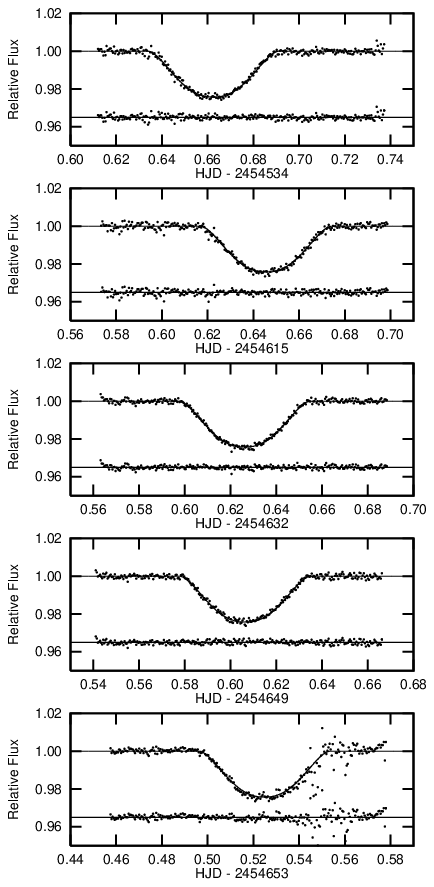}
\caption{RISE light curves of TrES-3 taken from 2008 March 8 to 2008 July 5 with their best fit models from the MCMC analysis over-plotted. Residuals from the best fit model are shown below each light curve offset.}
\label{fig:lightcurves1}
\end{figure}

\begin{figure}
\epsscale{.60} 
\plotone{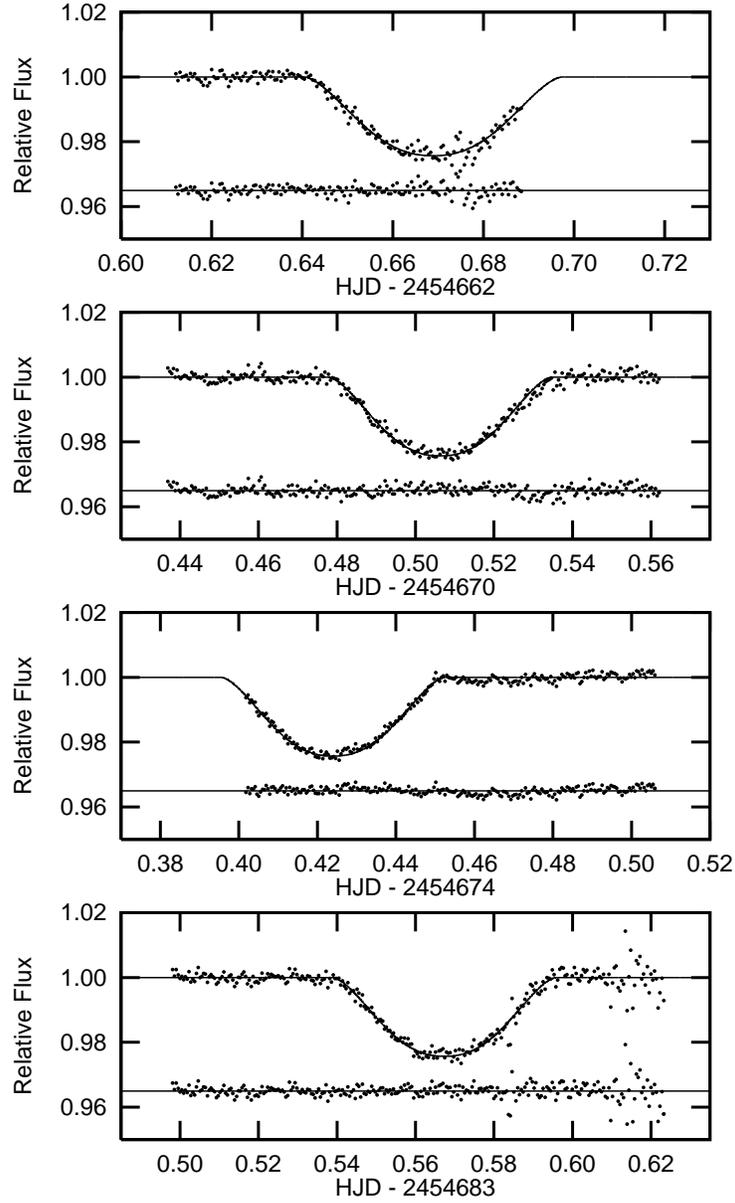}
\caption{Same as Figure~\ref{fig:lightcurves1}, for light curves from 2008 July 14 to 2008 August 4.}
\label{fig:lightcurves2}
\end{figure}

The flux of TrES-3 was then divided by the sum of the flux from the companion stars (all checked to be non-variable) to obtain each lightcurve. Initial estimates of the photometric errors were calculated using the aperture electron flux, sky and read noise. The lightcurves were then normalised by dividing through with a linear function of time fitted to the out-of-transit data, setting the unocculted flux of TRES-3 equal to 1. The light curves, along with their best fit models and residuals (see \S\ref{sect:system_parameters}), are shown in Figures~\ref{fig:lightcurves1} and \ref{fig:lightcurves2}.

\section{Light curve modelling and analysis}
\label{sect:modelling}
\subsection{Determination of system parameters}
\label{sect:system_parameters}

In order to determine the system parameters from the transit light curves, a parameterised model was constructed as in \citet{gibson_2008}. This used Kepler's Laws and assumed a circular orbit to calculate the normalised separation ($z$) of the planet and star centres as a function of time from the stellar mass and radius  ($M_{\star}$ and $R_{\star}$), the planetary mass and radius ($M_{p}$ and $R_{p}$), the orbital period and inclination ($P$ and $i$), and finally a central transit time for each lightcurve ($T_{0,n}$).  The analytic models of \citet{mandel_agol} were then used to calculate the stellar flux occulted by the planet from the normalised separation and the planet/star radius ratio ($\rho$) assuming the quadratic limb darkening function
$$
\frac{I_{\mu}}{I_1} = 1 - a(1-\mu) - b(1-\mu)^2,
$$
where $I$ is the intensity, $\mu$ is the cosine of the angle between the line-of-sight and the normal to the stellar surface, and $a$ and $b$ are the linear and quadratic limb darkening coefficients, respectively.

Limb darkening parameters were obtained from the models of \citet{claret}. We linearly interpolated the ATLAS tables for $T_{eff}$ = 5\,650\,K, log\,$g$ = 4.4, [Fe/H] = -0.19 and $v_{t}$ = 2.0\,kms$^{-1}$ \citep[from][]{sozzetti_2008} to obtain limb darkening parameters in both the V and R bands. The average from the V and R bands was then adopted as our theoretical limb darkening parameters. Several tests were performed to examine the effects of the choice of limb darkening parameters on the results, which are described at the end of this section.

A Markov-Chain Monte-Carlo (MCMC) algorithm was then used to obtain the best fit parameters and their uncertainties \citep[see e.g.,][]{tegmark, tlc1,  tlc9, cameron}.  This consists of calculating the $\chi^2$ fitting statistic,
$$
\chi^2=\sum_{j=1}^{N}\frac{(f_{j,obs} - f_{j,calc})^2}{\sigma^2_j}  
+ \frac{(M_{\star} - M_0)^2}{{\sigma_{M_0}}^2},
$$
where  $f_{j,obs}$ is the flux observed at time $j$, $\sigma_j$ is the corresponding uncertainty and  $f_{j,calc}$ is the flux calculated from the model for time $j$ and for the set of physical parameters described above. The second term represents a Gaussian prior placed on $M_{\star}$, where $M_0$ and 
$\sigma_{M_0}$ are the stellar mass and uncertainty as given in \citet{sozzetti_2008}. This allows the stellar mass to vary within constraints for each model fit, so that errors in the stellar mass are taken into account when extracting errors from the MCMC distributions. The stellar radius was updated for each choice of $M_{\star}$ using the scaling relation $R_{\star} \propto M_{\star}^{1/3}$, whilst $P$ and $M_{p}$ were held fixed at their previously determined values, as their uncertainties do not have any significant effect on the output probability distributions. Subsequent parameter sets are then chosen by perturbing small amounts to the previously accepted parameter set and are then accepted with probability $\exp(-\Delta\chi^2/2)$
at each point in the chain, where $\Delta\chi^2$ represents the difference in $\chi^2$ calculated for the old and new parameter sets. The procedure is the same as that used in \citet{gibson_2008}, to which the reader is referred for details.

To obtain reliable estimates of parameters and their uncertainties, it is important that the photometric errors are calculated accurately. The photometric errors $\sigma_j$ are first rescaled so that the best fitting model for each lightcurve has a reduced $\chi^2$ of 1. It is also vital to account for any correlated (``red'') noise in the data \citep[see e.g.,][]{pont2006, gillon2006}. The same procedure was used as in \citet{gibson_2008}, where we evaluated the presence of red noise in each light curve by calculating a factor $\beta~(\ge 1)$ according to \citet{tlc9} and rescaled the photometric errors by this value. A value for $\beta$ is determined by analysing the residuals from the best fit model of the lightcurves. Calculating the standard deviation of the residuals $\sigma_1$, and the standard deviation after binning the residuals into $M$ bins of $N$ points $\sigma_N$, one would expect
$$
\sigma_N = \frac{\sigma_1}{\sqrt{N}}\sqrt{\frac{M}{M - 1}}
$$
in the absence of red noise. This is usually larger by a factor $\beta$. However, the value determined for $\beta$ depends strongly on the choice of averaging time, i.e. $M$ and $N$. Previously we have used an average of $\beta$ values in the range 10--35 mins (the approximate time-scale of ingress or egress) to rescale the photometic errors. However, for this analysis we decided to use the maximum value for $\beta$ in this range in order to be as conservative as possible in determining our resultant errors.

Normalisation plays an important role in determining parameters and errors from light curves, and to account for this a further 2 parameters were added to the model for each transit.  These were the out-of-transit flux ($f_{oot,n}$) and a time gradient ($t_{Grad,n}$), which are vital for TTV measurements as these affect the symmetry of the light curve and therefore the central transit times. An airmass correction was not used as previous studies have shown this produces similar results for full transits, but impedes chain convergence for partial transits \citep{gibson_2008}.

An initial MCMC analysis was used to estimate the starting parameters and jump functions for $\rho$, $i$, $T_{0,n}$, $f_{oot,n}$ and $t_{Grad,n}$. An MCMC run was then started for all nine light curves, fixing the central transit times to those determined in the initial run. Other parameters, such as the normalisation parameters, that are independent for each light curve were still allowed to vary. Five separate chains with 200\,000 points were then computed with the initial \emph{free} parameters set by adding a $5\sigma$ gaussian random to their previously determined best fit values. The first 20\% of each chain was eliminated to keep the initial conditions from influencing the results, and the remaining parts of the chains were merged to obtain the best fit values and uncertainties for each free parameter. The best fit value was set as the modal value of the probability distribution, and the $1\sigma$ limits to the values where the integrals of the distribution from the minimum and maximum values were equal to 0.159. To test that the chains had all converged to the same region of parameter space, the Gelman \& Rubin statistic \citep{gelrub} was then calculated for each of the free parameters, and was found to be less than 0.5\% from unity for all parameters, a good sign of mixing and convergence.

To check for any errors that may have resulted from a poor choice of limb darkening parameters, the above procedure was repeated, this time allowing the linear limb darkening parameter ($a$) to vary freely whilst holding the quadratic limb darkening parameter ($b$) fixed at the theoretical value, as in \citet{southworth_2008}. This, however, results in unphysical models of the limb darkening (negative values), as was found in \citet{sozzetti_2008}, but not by \citet{gibson_2008} using the same technique for the WASP-3 system. This is probably due to the higher impact parameter of TRES-3, and therefore a higher sensitivity to the limb darkening parameters.  As a compromise, the same a priori constraint was imposed on the linear limb darkening as used in \citet{sozzetti_2008}, assuming that the limb darkening parameters do not drift from their theoretical values by more than 0.2 \citep{southworth_2008}. This involved adding another term to the $\chi^2$ function as follows;
$$
\chi^2=\sum_{j=1}^{N}\frac{( f_{j,obs} - f_{j,calc})^2}{\sigma^2_j}  
+ \frac{(M_{\star} - M_0)^2}{{\sigma_M}^2}
+ (\frac{a - a_0}{0.2})^2,
$$
where $a_0$ represents the theoretical limb darkening coefficient. As this causes significant increases in the errors determined for $i$ and $\rho$, these results were adopted as our final system parameters.

A further two checks were performed to test the limb darkening parameters. The first involved repeating this process by replacing the quadratic limb darkening coefficient determined for the combined V+R filter by that obtained for the individual V and R filters. This causes no significant changes to our results. The second check involved allowing each light curve to have its own independently varying (linear) limb darkening coefficient (within the prior constraints), rather than having one set of limb darkening coefficients to describe all the light curves. This again caused no significant changes to our results. It is therefore favourable to have the same set of limb darkening coefficients to describe all of the light curves, as forcing the same transit shape may reveal systematics through small differences in each light curve, that could otherwise be hidden through varying the limb darkening coefficients independently.


\subsection{Central Transit Times}

In order to calculate the central transit times for a TTV analysis we used the MCMC code, as described above, on each individual light curve, this time keeping the system parameters $\rho$ and $i$ fixed at the best fit values determined in the previous section. Modeling the light curves individually has the advantage of needing much shorter chains, and does not result in underestimated uncertainties. This is because the central transit times are not very sensitive to the physical system parameters, but rather to those parameters that effect the symmetry of the light curves, in particular the normalisation function. The same analysis was done on the light cuves from  \citet{sozzetti_2008}, so that the central transit times and errors were found using consistent methods.

For each light curve, five chains of length 50\,000 were computed, and $T_0$ and its uncertainty were extracted as before from the probability distribution after merging the chains (again discarding the first 20\% of each). The linear limb darkening coefficient, stellar mass, and stellar radius were allowed to vary within the same constraints outlined before. Again, the Gelman \& Rubin statistic was used to check for convergence. The systematics were accounted for using the same technique as described in the previous section, by re-scaling the errors of each light curve by a factor $\beta$. Values for $\beta$ are given in Tables~\ref{tab:RISE_timings} and \ref{tab:sozzetti_timings}.

A residual-permutation (RP) or ``prayer bead'' algorithm  \citep[see e.g.][]{southworth_2008, gillon_2008} was also used on each of the light curves to determine the errors in the transit times. This is another method commonly used to evaluate the effects of systematic noise on light curves, which often results in larger uncertainties than the MCMC method.
 
The RP method consists of reconstructing the light curve by adding the residuals to the best fit model from the MCMC fit, each time shifting the residuals by a random amount, and performing a new fit on the light curve. 10\,000 such fits were performed for each transit, with $M_{\star}$, $R_{\star}$, $i$ and  $\rho$ selected from a Gaussian distribution at the start of each using the stellar parameters and uncertainties from \citet{sozzetti_2008}, and the system parameters determined for the combined RISE light curves. The transit times, normalisation parameters, and linear limb darkening co-efficient were allowed to vary freely, using starting points determined randomly within 10$\sigma$ from the best fit values. Errors in the central transit times were then estimated from the resulting distribution of fits.

This method has the advantage that it preserves the actual correlated noise from the light curve, whereas the error re-scaling technique used alongside the MCMC fitting is sensitive to the choice of averaging time. However, a comparison of the two methods showed, that in most cases, the errors from the MCMC fit were larger than those from RP. This is likely due to choosing the maximum value for $\beta$ in the 10--35\,min range to re-scale the photometric errors prior to the MCMC runs, rather than using the average value. 

For each transit the ``worst case'' was assumed, i.e. we adopted the error from the RP method only when it produced a larger uncertainty than the MCMC code. Note we always used best fit values from the MCMC fit as the RP method already assumed these transit times when reconstructing each light curve. The methods used to determine each of the timing errors are given in Tables~\ref{tab:RISE_timings} and \ref{tab:sozzetti_timings}. 


\section{Results}
\label{sect:results}
\subsection{System parameters}

The system parameters derived from the MCMC fits of the RISE transits are shown in Table~\ref{tab:system_parameters}. \citet{sozzetti_2008} undertook a thorough analysis of the stellar properties and radial velocities, and therefore we focus only on the planet parameters that are observable in the light curves, namely the inclination of the orbit and the ratio of the planet to stellar radius. We found $i = 81.73^{+0.13}_{-0.04}$, and $\rho = 0.1664^{+0.0011}_{-0.0018}$. These are consistent with previously determined values, although with slightly smaller uncertainties, and therefore the planet radius and density \citep[derived from the stellar radius and planetary mass from][]{sozzetti_2008} are also consistent with previous studies.


\subsection{Transit ephemeris}

The central transit times are shown in Table~\ref{tab:RISE_timings} for the RISE light curves, and in Table~\ref{tab:sozzetti_timings} for the light curves of \citet{sozzetti_2008}, where the transit times were found to be consistent to within $\sim0.3\sigma$ in all cases, and typically to less than $0.1 \sigma$. 
Due to the more rigourous approach used to account for red noise in our analysis, the error bars were found to be $\sim10-40\%$ larger.

\begin{figure*}
\epsscale{0.80} 
\plotone{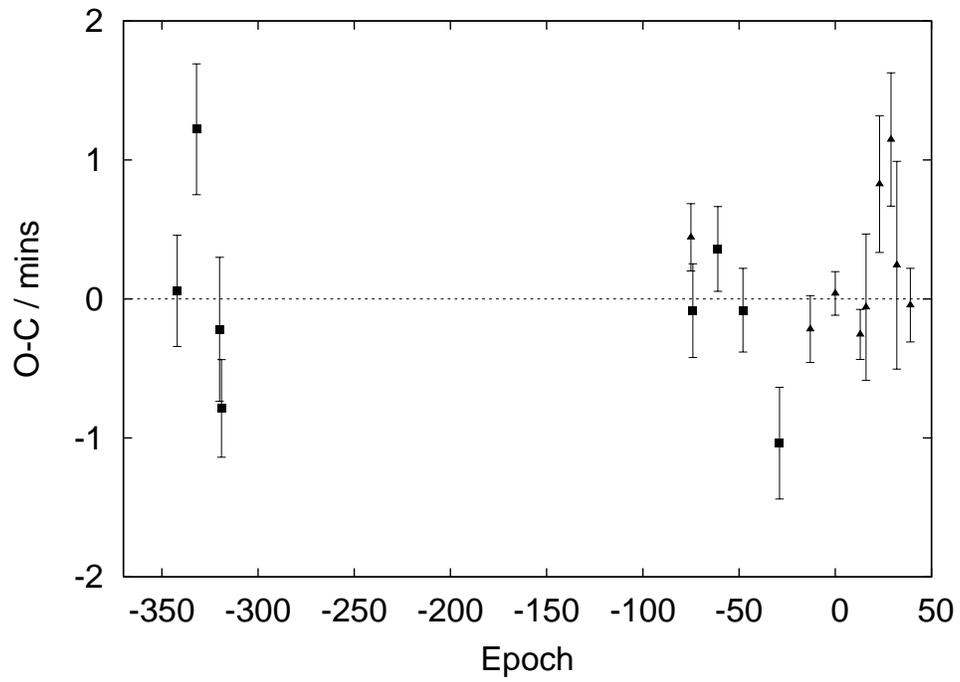}
\caption{Timing residuals of the RISE transits (triangles) and those from \citet[][squares]{sozzetti_2008}.} 
\label{fig:ttvplot} 
\end{figure*}

A new ephemeris was calculated by minimising $\chi^2$ through fitting a linear function of Epoch $E$ and Period $P$ to the transit times
$$
T_c(E) = T_c(0) + EP,
$$
where $E = 0$ was set to the transit from 2008 June 14 taken with RISE, as it has the smallest uncertainty. The results were $T_c(0) = 2454632.62610 \pm 0.00006$ and $P =1.3061864 \pm 0.0000005$. Figure~\ref{fig:ttvplot} shows a plot of the timing residuals of the RISE and \citet{sozzetti_2008} transits using this updated ephemeris.

For the RISE data a straight line fit yeilds $\chi^2 = 13.49$ for 7 degrees of freedom, and for the \citet{sozzetti_2008} data gives $\chi^2 = 19.40$ for 6 degrees of freedom, much lower than the value of 35.22 found from their analysis, simply because of the larger timing errors. For the combined data set, $\chi^2 = 35.07$ for 15 degrees of freedom, and therefore a reduced $\chi^2$ of 2.34. This all seems to support the conclusions from \citet{sozzetti_2008}, that the uncertainties are underestimated, or that a linear period is not a good fit to the data. As we have been as sceptical as possible regarding the timing errors, this seems to suggest tentative evidence of a third body in the system perturbing the orbit of TrES-3b.

However, after closer inspection of the light curves that contribute most to $\chi^2$, this conclusion is less convincing. Most of the large contributors to $\chi^2$ are partial transits or have very little out-of-transit data. If we remove all of the light curves with less than 20 mins of out-of-transit data either before ingress or after egress (transits $E$ = -332, -319, -29, 23 and 32) this results in a $\chi^2$ of 13.53 for 10 degrees of freedom, or a reduced $\chi^2$ of 1.35. This is because these transits are much more difficult to normalize due to lack of out-of-transit data, and unseen systematics could certainly cause the normalization gradient to be skewed, therefore effecting the symmetry of the light curves and hence the central transit times. This seems to suggest that transits need at least $\sim20-30$ mins of out-of-transit data either side of the transit to be useful for transit timing studies, unless a more robust method of normalizing light curves and accounting for the errors is found. The largest remaining contributor to $\chi^2$ is transit $E$ = 29 which lies $\sim2.4\sigma$ from the straight line fit. This transit not only has a high level of red-noise ($\beta>2$), but a dip in the residuals from the best fit model is clearly seen around egress. The net effect of this on the model fit would be to ``drag'' the measurement of the central transit time later, as seen in the transit timing residuals. Removing this transit results in a reduced $\chi^2 < 1$, which suggests a constant period. Conclusions supporting a third body in this system would therefore rely on transits with little out-of-transit coverage and/or those with large visible systematics.

\subsection{Limits on a second planet in the TrES-3 system.}
\label{sect:upper_mass_limits}

Despite not revealing a significant TTV signal, the data can still be used to place upper mass limits on the presence of a hypothetical second planet in the TrES-3 system. The shape and amplitude of transit timing residuals are dependent on a large number of parameters, such as mass, period, eccentricity and argument of periastron of the perturbing planet.

In order to compute model timing residuals, the equations of motion for a three body system were integrated using a 4th order Runge-Kutta method. The first two bodies were set to represent the star and planet of the TrES-3 system, which was assumed to have an initially circular orbit. Transit times were then extracted when the star and the transiting planet were aligned along the direction of observation, with the third body representing the perturbing planet. The transit times were then fit with a linear function of time and the timing residuals used for comparison with the data. The orbits of the planets and direction of observation were assumed to be coplanar.

Ideally we would like to search the parameter space of the perturbing planet completely and set upper mass limits at each point. However, this is not possible given the large amount of computation required to produce a model of timing residuals at each point in such a large parameter space. Therefore, some simplifications and assumptions were made. Firstly, we assumed that the amplitude of the timing residuals for a given perturbing orbital configuration are proportional to the mass of the perturbing planet \citep{agol2005, holman_murray}, and verified this by constructing models with a range of perturbing planet masses. Secondly, we assumed that the perturbing planet had a starting eccentricity of 0, as an increase in eccentricity generally increases the amplitude of the timing residuals and therefore to set upper masses as a function of period we only need to investigate perturbing planets on circular orbits. This assumption is tested later in this section.

Models were created for an Earth-massed perturbing planet with a period ratio distributed from 0.2 to 5.0 (the regime in which relatively small masses may be detected), increasing the sampling around both the interior and exterior 2:1 resonances, where we expect to probe for the smallest masses. For each model produced, the transit times were extracted for a range of observation directions. 

To calculate the maximum allowed mass for each model, $\chi^2$ was calculated by fitting the model residuals to the measured timing residuals from the light curves. The mass of the perturbing planet was increased (or decreased) by scaling the timing residuals until $\chi^2$ was increased by a value $\Delta\chi^2 = 9$ \citep{steffen_agol2005, agol_2007} from that of a constant period (ie timing residuals = 0) which corresponds to a $3\sigma$ confidence limit. We then minimised $\chi^2$ along epoch only, and then let the mass of the perturber grow again until the maximum allowed mass for each model was determined. This procedure was repeated for the range of observation directions of each model, and the largest upper mass determined was assumed as our upper mass limit for each period ratio.

To check that the starting mass of each model had no impact on the mass limits found for each period (i.e. test that residuals are indeed proportional to the perturbing mass), models were re-calculated with the mass of the perturbing planet set as the upper mass limits found from the $\chi^2$ fits. Upper masses limits were then determined as before. This process was repeated twice and was found to make little difference to the final upper mass limits, therefore justifying our assumption. 

\begin{figure*}
\epsscale{1.0} 
\plotone{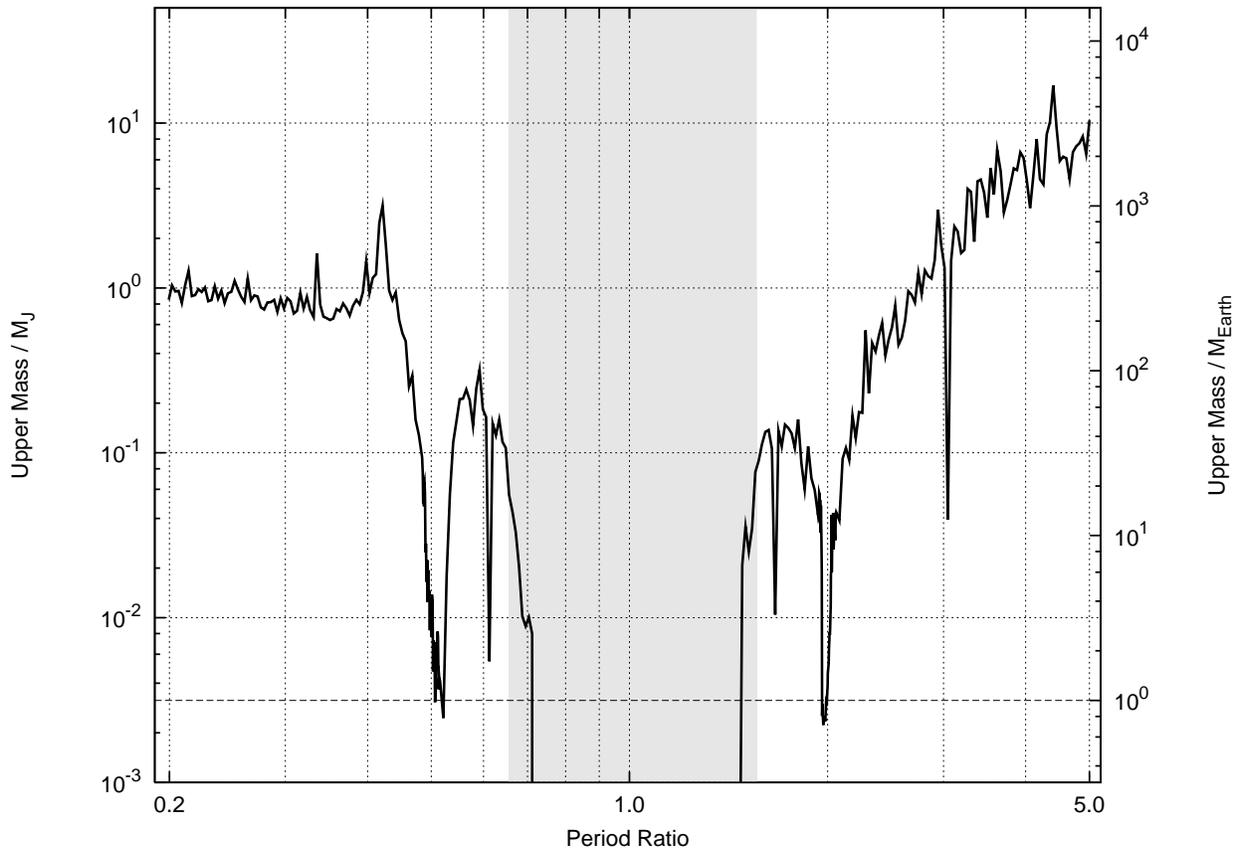}
\caption{Upper mass limits of a hypothetical 2nd planet in the TrES-3 system as a function of period ratio. The solid black line represents the upper mass found for the 3 body simulations, and the horizontal dashed line represents an Earth-mass planet. The region where an Earth-massed planet is not guaranteed to be Hill stable is marked by the grey shading.}
\label{fig:upper_mass_plot} 
\end{figure*}

Figure~\ref{fig:upper_mass_plot} shows a plot of the resulting upper mass limits found as a function of the period ratio. The solid black line shows the upper mass limits found for the 3 body simulations, and the horizontal dashed line represents an Earth-mass planet. The results show that we have probed for masses as low as 0.97 $M_{\oplus}$ and 0.71 $M_{\oplus}$ in the interior and exterior 2:1 resonances, respectively. 

To test our assumption that perturbers on initially circular orbits will cause the smallest perturbations, and therefore can be used to set upper mass limits on a perturbing mass as a function of period ratio, we created models this time allowing the perturbing mass to have non-zero initial eccentricity.
A set of models was created with a period range spanning the exterior 2:1 resonance and the perturbing bodies eccentricity ranging from 0 to 0.15. It was found that generally the amplitude of the signal drops and reaches a minimum between $e\sim$0.01 and 0.12, before increasing again, and could drop by as much as an order of magnitude. A similar (but smaller) effect was found for the interior 2:1 resonance. This invalidates our assumption, and suggests that to set upper mass limits around resonance at least the period, eccentricity and argument of periastron of the perturbing planet needs to be explored in parameter space, which is beyond the scope of this paper.
However, upper mass limits were estimated using this set of models and the same technique as before, and we found that more realistic upper mass limits are $\sim3-4M_{\oplus}$ and $\sim10-15M_{\oplus}$ in the interior and exterior 2:1 resonances, respectively. Out of resonance, it was found that the amplitude of TTV signals increases with eccentricity of the perturbing planet, and thus our assumption and upper mass limits are valid.

A long term stability analysis was not performed for the 3-body systems. \citet{bean_2009} found that for the CoRoT-1 system, only test particles with period ratios greater than $\sim1.8$ were stable for more than $10^6$ orbits of the transiting planet. CoRoT-1 has a similar G-dwarf host star and a slightly longer period ($\sim1.5$ days), which suggests that a similar analysis would prove useful here. \citet{barnes_greenberg_2006} explore the stability limits in exoplanet systems, and provide an inequality to test whether a system is Hill stable (equation 2). Using this inequality for the TrES-3 system (assuming an Earth-massed perturber), places lower and upper limits on the period ratio of 0.64 and 1.59, respectively. The resulting region not guaranteed to be Hill stable is marked on figure~\ref{fig:upper_mass_plot} by the grey shading, although stable configurations may still occur in this region. Trojan companions could also exist in stable orbits near the 1:1 resonance. \citet{madhusudan_winn_2009} placed a $2\sigma$ upper mass limit of 81.3 Earth masses on a trojan in the TrES-3 system by combining transit observations and radial velocity data.

\section{Summary and discussion}
\label{sect:summary}
This paper presents the first transits taken using RISE specifically for a transit timing analysis, consisting of nine light curves of TrES-3. The transits were fit with an MCMC code and the derived system parameters found to be consistent with previous studies. Two different methods were used to determine the errors in the  central transit times, trying to take into account the systematics in the light curves. These were scaling the errors bars by a constant prior to MCMC fitting after analysing the residuals, and using a residual permutation algorithm. The largest error found was used for each transit. We have shown that when systematics are kept at a minimum, it is possible to determine transit times to $\sim$10 seconds - the level of accuracy expected from RISE.

Whilst the transit times appear to deviate significantly from a constant period when a $\chi^2$ analysis is performed, those that contribute most to the deviations tend to have very little out of transit data or obvious systematics. After removing transits with less than $\sim$20 mins of out-of-transit coverage either before ingress or after egress, the data are consistent with a constant period, and therefore no conclusive evidence was found for the presence of a second planet in the TrES-3 system. The transit times were then used to place upper mass limits on a perturbing planet as a function of period that could be present in the system yet not detected through our observations. This showed that our observations were sensitive to Earth-mass planets or smaller in the interior and exterior 2:1 resonances when we assume the perturbing mass is on an initially circular orbit. However, larger planets may exist in low eccentricity orbits around the 2:1 resonances, and exploring period, eccentricity and argument of periastron in parameter space is called for to set true upper mass limits, as well as a long term stability analysis.

This study highlights the difficulties in attempting to detect a TTV signal, as we need to have complete confidence in the error bars calculated, which may require a more robust method to normalize the light curves and deal with red noise, especially if we are to trust transits with limited out-of-transit data. Confirming a true TTV signal may therefore require some obvious structure in the observed residuals (which we would expect for resonant systems), rather than relying on a larger than expected ``scatter'' of points.

\acknowledgments
RISE was designed and built with resources made available from Queens University Belfast, Liverpool John Moores University and the University of Manchester. The Liverpool Telescope is operated on the island of La Palma by Liverpool John Moores University in the Spanish Observatorio del Roque de los Muchachos of the Instituto de Astrofisica de Canarias with financial support from the UK Science and Technology Facilities Council. D.L.P. was supported by a Leverhulme Research Fellowship for the duration of this work. F.P.K. is grateful to AWE Aldermaston for the award of a William Penney Fellowship. We also thank A. Sozzetti, for making his data available for re-analysis, and the referee, D. Fabrycky, for comments which improved the content of this paper.



{\it Facilities:} \facility{Liverpool:2m (RISE)}




\clearpage

\clearpage 

\clearpage

\clearpage


\clearpage
\begin{table*}
\begin{center}
\caption{Summary of the RISE light curves of TrES-3.}
\label{tab:lightcurves}
\begin{tabular}{lcccc}
\tableline\tableline
 Night & No. exposures & No. comparison & Aperture size & RMS (residuals)\\
 ~ & ~ & stars & (pixels) & (mmag)\\
 \tableline
2008 Mar 8 & 1350 & 4 & 8 & 1.31\\
2008 May 28 & 1350 & 4 & 7 & 1.42\\
2008 Jun 14 & 1350 & 7 & 5 & 1.00\\
2008 Jul 1 & 1350 & 6 & 4 & 1.10\\
2008 Jul 5 & 1350 & 7 & 6 & 3.31\\
2008 Jul 14 & 825   & 6 & 7 & 1.81\\
2008 Jul 22 & 1350 & 2 & 4 & 1.57\\
2008 Jul 26 & 1125 & 8 & 4 & 1.18\\
2008 Aug 4 & 1350 & 2 & 4 & 2.60\\
\tableline
\end{tabular}
\end{center}
\end{table*}

\begin{table*}
\begin{center}
\caption{Parameters and $1\sigma$ uncertainties for TrES-3 as derived from MCMC fitting of RISE light curves and some further calculated parameters.}
\label{tab:system_parameters}
\begin{tabular}{lccc}
 
\tableline\tableline
 Parameter & Symbol & Value & Units\\
\tableline
Planet/Star radius ratio	&	$\rho$         	& $0.1664^{+0.0011}_{-0.0018}$ 				&\\
Orbital inclination 		&	$i$  			& $81.73^{+0.13}_{-0.04}$		     			& $\deg$\\
Impact parameter		&	$b$  			& $0.852^{+0.004}_{-0.013}$ 					&\\ 
Transit duration		&	$T_d$		& $1.332^{+0.024}_{-0.010}$ 					& hours\\
Transit epoch			&	$T_0$		& $2454632.62610 \pm 0.00006$ 				&HJD\\
Period				&	$P$			& $1.3061864 \pm 0.0000005$				& days\\
Planet radius			&	$R_p$         	& $1.341^{+0.025}_{-0.035}$ 					& $R_J$\\
Planet mass\tablenotemark{a}	&	$M_p$		& $1.910^{+0.075}_{-0.080}$  					& $M_J$\\ 
Planet density			&	${\rho}_p$  	& $0.792^{+0.047}_{-0.042}$					& ${\rho}_J$\\ 
Planetary surface gravity	& 	log $g_p$		& $3.421^{+0.023}_{-0.022}$                                         & [cgs]\\
\tableline
\end{tabular}
\tablenotetext{a}{From \citet{sozzetti_2008}, displayed here for convenience.}
\end{center}
\end{table*}

\clearpage
\begin{table*}
\begin{center}
\caption{Central transit times and uncertainties for the RISE photometry including the error source.}
\label{tab:RISE_timings}
\begin{tabular}{lcccc}
\tableline\tableline
Epoch & Central Transit Time & Uncertaintly & $\beta$\tablenotemark{b} & Error Source\\
~ & [HJD] & (days) & ~  & ~ \\ 
\tableline
-75  & 2454534.66243 & 0.00017 & 1.07 & RP \\
-13 & 2454615.64553 & 0.00017  & 1.49 & MCMC \\
~0 & 2454632.62613 & 0.00011   & 1.25 & MCMC \\
~13 & 2454649.60634 & 0.00013 & 1.31 & MCMC \\
~16 & 2454653.52504 & 0.00037 & 1.95 & MCMC \\
~23 & 2454662.66896 & 0.00034 & 1.56 & MCMC \\
~29 & 2454670.50630 & 0.00033 & 2.04 & RP \\
~32 & 2454674.42423 & 0.00052 & 2.83 & RP \\
~39 & 2454683.56734 & 0.00018 & 1.00 & MCMC \\
\tableline
\end{tabular}
\tablenotetext{b}{Re-scale factor from red noise analysis (See \S\ref{sect:system_parameters}).}
\end{center}
\end{table*}

\begin{table*}
\begin{center}
\caption{Central transit times and uncertainties for the light curves of \citet{sozzetti_2008} including the error source.}
\label{tab:sozzetti_timings}
\begin{tabular}{lcccc}
\tableline\tableline
Epoch & Central Transit Time & Uncertaintly & $\beta$\tablenotemark{c} & Error Source\\
~ & [HJD] & (days) & ~  & ~ \\ 
\tableline
-342 & 2454185.91040 & 0.00028 & 1.63 & MCMC \\
-332 & 2454198.97307 & 0.00033 & 1.52 & MCMC \\
-320 & 2454214.64631 & 0.00036 & 1.58 & MCMC \\
-319 & 2454215.95210 & 0.00024 & 1.29 & MCMC \\
-74 & 2454535.96825 & 0.00023   & 1.39 & MCMC \\
-61 & 2454552.94898 & 0.00021  & 1.52 & MCMC \\
-48 & 2454569.92910 & 0.00021  & 1.34 & MCMC \\
-29 & 2454594.74597 & 0.00028  & 1.30 & MCMC \\
\tableline
\end{tabular}
\tablenotetext{c}{See Table~\ref{tab:RISE_timings}.}
\end{center}
\end{table*}


\begin{thebibliography}{26}
\expandafter\ifx\csname natexlab\endcsname\relax\def\natexlab#1{#1}\fi

\bibitem[{{Agol} {et~al.}(2005){Agol}, {Steffen}, {Sari}, \&
  {Clarkson}}]{agol2005}
{Agol}, E., {Steffen}, J., {Sari}, R., \& {Clarkson}, W. 2005, \mnras, 359, 567

\bibitem[{{Agol} \& {Steffen}(2007)}]{agol_2007}
{Agol}, E. \& {Steffen}, J.~H. 2007, \mnras, 374, 941

\bibitem[{{Barnes} \& {Greenberg}(2006)}]{barnes_greenberg_2006}
{Barnes}, R. \& {Greenberg}, R. 2006, \apjl, 647, L163

\bibitem[{{Bean}(2009)}]{bean_2009}
{Bean}, J.~L. 2009, ArXiv e-prints

\bibitem[{{Claret}(2000)}]{claret}
{Claret}, A. 2000, \aap, 363, 1081

\bibitem[{{Collier Cameron} {et~al.}(2007){Collier Cameron}, {Wilson}, {West},
  {Hebb}, {Wang}, {Aigrain}, {Bouchy}, {Christian}, {Clarkson}, {Enoch},
  {Esposito}, {Guenther}, {Haswell}, {H{\'e}brard}, {Hellier}, {Horne},
  {Irwin}, {Kane}, {Loeillet}, {Lister}, {Maxted}, {Mayor}, {Moutou}, {Parley},
  {Pollacco}, {Pont}, {Queloz}, {Ryans}, {Skillen}, {Street}, {Udry}, \&
  {Wheatley}}]{cameron}
{Collier Cameron}, A., {Wilson}, D.~M., {West}, R.~G., {et~al.} 2007, \mnras,
  380, 1230

\bibitem[{{Ford} \& {Gaudi}(2006)}]{ford_gaudi_2006}
{Ford}, E.~B. \& {Gaudi}, B.~S. 2006, \apjl, 652, L137

\bibitem[{{Ford} \& {Holman}(2007)}]{ford_holman_2007}
{Ford}, E.~B. \& {Holman}, M.~J. 2007, \apjl, 664, L51

\bibitem[{{Gelman} \& {Rubin}(1992)}]{gelrub}
{Gelman}, A. \& {Rubin}, D.~B. 1992, Stat. Sci., 7, 457

\bibitem[{{Gibson} {et~al.}(2008){Gibson}, {Pollacco}, {Simpson}, {Joshi},
  {Todd}, {Benn}, {Christian}, {Hrudkov{\'a}}, {Keenan}, {Meaburn}, {Skillen},
  \& {Steele}}]{gibson_2008}
{Gibson}, N.~P., {Pollacco}, D., {Simpson}, E.~K., {et~al.} 2008, \aap, 492,
  603

\bibitem[{{Gillon} {et~al.}(2006){Gillon}, {Pont}, {Moutou}, {Bouchy},
  {Courbin}, {Sohy}, \& {Magain}}]{gillon2006}
{Gillon}, M., {Pont}, F., {Moutou}, C., {et~al.} 2006, \aap, 459, 249

\bibitem[{{Gillon} {et~al.}(2009){Gillon}, {Smalley}, {Hebb}, {Anderson},
  {Triaud}, {Hellier}, {Maxted}, {Queloz}, \& {Wilson}}]{gillon_2008}
{Gillon}, M., {Smalley}, B., {Hebb}, L., {et~al.} 2009, \aap, 496, 259

\bibitem[{{Heyl} \& {Gladman}(2007)}]{heyl_gladman}
{Heyl}, J.~S. \& {Gladman}, B.~J. 2007, \mnras, 377, 1511

\bibitem[{{Holman} \& {Murray}(2005)}]{holman_murray}
{Holman}, M.~J. \& {Murray}, N.~W. 2005, Science, 307, 1288

\bibitem[{{Holman} {et~al.}(2006){Holman}, {Winn}, {Latham}, {O'Donovan},
  {Charbonneau}, {Bakos}, {Esquerdo}, {Hergenrother}, {Everett}, \&
  {P{\'a}l}}]{tlc1}
{Holman}, M.~J., {Winn}, J.~N., {Latham}, D.~W., {et~al.} 2006, \apj, 652, 1715

\bibitem[{{Kipping}(2009)}]{kipping_2009}
{Kipping}, D.~M. 2009, \mnras, 392, 181

\bibitem[{{Madhusudhan} \& {Winn}(2009)}]{madhusudan_winn_2009}
{Madhusudhan}, N. \& {Winn}, J.~N. 2009, \apj, 693, 784

\bibitem[{{Mandel} \& {Agol}(2002)}]{mandel_agol}
{Mandel}, K. \& {Agol}, E. 2002, \apjl, 580, L171

\bibitem[{{Miralda-Escud{\'e}}(2002)}]{miralda}
{Miralda-Escud{\'e}}, J. 2002, \apj, 564, 1019

\bibitem[{{O'Donovan} {et~al.}(2007){O'Donovan}, {Charbonneau}, {Bakos},
  {Mandushev}, {Dunham}, {Brown}, {Latham}, {Torres}, {Sozzetti}, {Kov{\'a}cs},
  {Everett}, {Baliber}, {Hidas}, {Esquerdo}, {Rabus}, {Deeg}, {Belmonte},
  {Hillenbrand}, \& {Stefanik}}]{ODonovan_2007}
{O'Donovan}, F.~T., {Charbonneau}, D., {Bakos}, G.~{\'A}., {et~al.} 2007,
  \apjl, 663, L37

\bibitem[{{Pont} {et~al.}(2006){Pont}, {Zucker}, \& {Queloz}}]{pont2006}
{Pont}, F., {Zucker}, S., \& {Queloz}, D. 2006, \mnras, 373, 231

\bibitem[{{Southworth}(2008)}]{southworth_2008}
{Southworth}, J. 2008, \mnras, 386, 1644

\bibitem[{{Sozzetti} {et~al.}(2009){Sozzetti}, {Torres}, {Charbonneau}, {Winn},
  {Korzennik}, {Holman}, {Latham}, {Laird}, {Fernandez}, {O'Donovan},
  {Mandushev}, {Dunham}, {Everett}, {Esquerdo}, {Rabus}, {Belmonte}, {Deeg},
  {Brown}, {Hidas}, \& {Baliber}}]{sozzetti_2008}
{Sozzetti}, A., {Torres}, G., {Charbonneau}, D., {et~al.} 2009, \apj, 691, 1145

\bibitem[{{Steele} {et~al.}(2008){Steele}, {Bates}, {Gibson}, {Keenan},
  {Meaburn}, {Mottram}, {Pollacco}, \& {Todd}}]{steele_2008}
{Steele}, I.~A., {Bates}, S.~D., {Gibson}, N., {et~al.} 2008, Proc. SPIE, 7014

\bibitem[{{Steffen} \& {Agol}(2005)}]{steffen_agol2005}
{Steffen}, J.~H. \& {Agol}, E. 2005, \mnras, 364, L96

\bibitem[{{Tegmark} {et~al.}(2004){Tegmark}, {Strauss}, {Blanton}, {Abazajian},
  {Dodelson}, {Sandvik}, {Wang}, {Weinberg}, {Zehavi}, {Bahcall}, {Hoyle},
  {Schlegel}, {Scoccimarro}, {Vogeley}, {Berlind}, {Budavari}, {Connolly},
  {Eisenstein}, {Finkbeiner}, {Frieman}, {Gunn}, {Hui}, {Jain}, {Johnston},
  {Kent}, {Lin}, {Nakajima}, {Nichol}, {Ostriker}, {Pope}, {Scranton},
  {Seljak}, {Sheth}, {Stebbins}, {Szalay}, {Szapudi}, {Xu}, {Annis},
  {Brinkmann}, {Burles}, {Castander}, {Csabai}, {Loveday}, {Doi}, {Fukugita},
  {Gillespie}, {Hennessy}, {Hogg}, {Ivezi{\'c}}, {Knapp}, {Lamb}, {Lee},
  {Lupton}, {McKay}, {Kunszt}, {Munn}, {O'Connell}, {Peoples}, {Pier},
  {Richmond}, {Rockosi}, {Schneider}, {Stoughton}, {Tucker}, {vanden Berk},
  {Yanny}, \& {York}}]{tegmark}
{Tegmark}, M., {Strauss}, M.~A., {Blanton}, M.~R., {et~al.} 2004, \prd, 69,
  103501

\bibitem[{{Winn} {et~al.}(2008){Winn}, {Holman}, {Torres}, {McCullough},
  {Johns-Krull}, {Latham}, {Shporer}, {Mazeh}, {Garcia-Melendo}, {Foote},
  {Esquerdo}, \& {Everett}}]{tlc9}
{Winn}, J.~N., {Holman}, M.~J., {Torres}, G., {et~al.} 2008, \apj, 683, 1076

\end{thebibliography}
\end{document}